\documentclass[aip,jcp,a4paper,reprint,nobibnotes,superscriptaddress]{revtex4-1}
\usepackage{amssymb,amsmath}
\usepackage{graphicx}
\usepackage{color}
\usepackage{microtype}
\usepackage{threeparttable}
\usepackage{soul}
\usepackage{multirow}
\usepackage{dcolumn}
\begin{document}

\title{Correlated Electronic Properties of a Graphene Nanoflake: Coronene}

\date{\today}

\author{Suryoday Prodhan}
\thanks{Current address: Laboratory for Chemistry of Novel Materials, University of Mons, Mons-7000, Belgium.}
\email[Electronic mail:]{suryodayp@iisc.ac.in}
\affiliation{Solid State and Structural Chemistry Unit, Indian Institute of Science, Bangalore-560012, India.}
\author{Sumit Mazumdar}
\email[Electronic mail:]{sumit@physics.arizona.edu}
\affiliation{Department of Physics, University of Arizona, Tucson, Arizona 85721, USA.}
\affiliation{College of Optical Sciences, University of Arizona, Tucson, Arizona 85721, USA.}
\author{S. Ramasesha}
\email[Electronic mail:]{ramasesh@iisc.ac.in}
\affiliation{Solid State and Structural Chemistry Unit, Indian Institute of Science, Bangalore-560012, India.}

\begin{abstract}
We report studies of the correlated excited states of coronene and substituted coronene
within the Pariser-Parr-Pople (PPP) correlated $\pi$-electron model employing symmetry
adapted density matrix renormalization group technique. These polynuclear aromatic hydrocarbons can be considered as graphene nanoflakes.
We review their electronic structures utilizing a new symmetry adaptation scheme that exploits
electron-hole symmetry, spin-inversion symmetry and end-to-end interchange symmetry. 
Study of the electronic structures sheds light on the electron correlation effects in 
these finite-size graphene analogues, which diminishes on going from one-dimensional 
to higher-dimensional systems, yet is significant within these finite graphene derivatives.
\end{abstract}


\maketitle

\section{Introduction}

In quantum chemical calculations of electronic structures of carbon (C)-based $\pi$-conjugated systems, 
the effects of electronic correlation have been probed by employing a number of techniques, 
most of which are variants of the restricted configuration interaction technique. 
Singles configuration interaction (SCI) based techniques suffice for studying one-photon optical gaps in neutral carbon-based 
molecular systems, as these excitations are predominantly single electron-hole excitations \cite{Mazumdar04}.
However, for strongly correlated low-dimensional systems, the shortcomings of the SCI
are also well-documented \cite{Hudson82,Soos84,Schulten79-1,Schulten79-2,Schulten87,Schulten86,Soos93,Baeriswyl,Shuai2000}.
Time-dependent density functional theory \cite{Watanabe06,Wong09,Malloci11} or GW approximation accompanied by Bethe-Salpeter 
correction \cite{Spataru04,Chang04,Louie09,Louie06-2} are essentially equivalent to the 
SCI approximation as only two-particle (one electron - one hole) interactions are 
included in these approaches and higher-order CIs are excluded. 

While full CI (FCI) study is the most preferred one, its 
use has been limited to $\sim 18$-electron neutral systems, as the Hilbert space 
dimension increases exponentially with system size. Ab initio quantum 
chemical methods like CASPT2 that are able to reproduce the 
energy spectra of correlated $\pi$-electron molecules correctly
are limited to $8-10$-electron neutral systems 
\cite{Thiel08,Thiel10,Dreuw06,Dreuw12,Dreuw13}. Multiple reference singles and doubles CI (MRSDCI) approach 
is another suitable iterative method for introduction of higher-order CI effects but the Hilbert space 
dimensions for different excited states, for similar accuracy, vary significantly \cite{Schulten87,Schulten86,Mazumdar14-1,
Mazumdar14-2}. On the other hand, the Density Matrix Renormalization Group (DMRG) method, introduced by White, 
is an accurate numerical many-body 
technique for studying low-lying states of one and quasi-one dimensional systems in real space 
\cite{White92,White93,Schollwock,Hallberg,Ramasesha}. In the DMRG method, similar to other renormalization group 
methods, the Hilbert space dimension remains fixed independent of the system sizes. For discrete molecular
systems with energy gaps, the area 
law of entanglement entropy also holds leading to high accuracy in the DMRG calculations 
\cite{Schollwock,Hallberg}, even with moderate dimensionality of block state space.   

The DMRG scheme, as usually implemented, utilizes conservation of the number of particles and the $z$-component 
of the total spin $S_z$. Thus, in this scheme, a few low-lying states are obtained in each of the 
sectors with a fixed number of particles and fixed $S_z$ value. Although schemes that exploit other symmetries 
have been developed, they are limited to a few orbitals and a few particles. Yet, in studies of $\pi$-conjugated systems
that probe the lowest-energy dipole-connected state, the lowest two-photon state and the lowest triplet state, we need to 
exploit other symmetries like the electron-hole ($e$-$h$) symmetry, the spin-inversion symmetry and the
end-to-end interchange symmetry (see below). 
However, although application of the DMRG technique or its symmetrized 
versions is straightforward in one- and quasi-one dimensional systems, it is not so trivial in higher 
dimensional systems like in graphene nanoflakes and graphene nanoribbons.
Study of these large finite graphene analogues 
is expected to shed light on the properties in the thermodynamic limit and the effect of
electronic correlation in these industrially important two-dimensional systems. Most of the earlier 
studies on graphene and graphene analogues have been done within non-interacting model or employing 
restricted configuration interaction technique with a few frontier molecular orbitals \cite{Wallace47,Macdonald07,Yu08,Gava09,
Giovannetti07,Shen08,Ribiero08,Latil06,Feng09,Fazzio07,Xu10,Pati14,Furst09,Tachikawa09,Scuseria11,Wang08,Wang09,Ezawa07,Palacios07,
Sheng13-1,Sheng13-2,Zhou13}, while the importance of electron correlation in the electronic and magnetic structures of these systems have been 
emphasized in recent studies \cite{Pati08,Mazumdar14-1,Mazumdar14-2,Goli16}.
In the present paper, we demonstrate the application of symmetrized DMRG technique in the
study of a graphene nanoflake, coronene, within a long-range correlated $\pi$-electron 
model. This molecule has recently been studied employing 
the MRSDCI approach \cite{Mazumdar14-1,Mazumdar14-2,Molinari14} and we 
reexamine the earlier results. We also study the effect of weak donor-acceptor substitutions which lowers the symmetry of the overall molecule. 
Transition dipole moments to the low-lying
optical states along with two-photon absorption cross-sections for the low-lying two-photon states are also calculated.

The paper is organized as follows. In the next section, we have given an account of the model Hamiltonian 
employed in our study along with a brief discussion about the DMRG technique and the symmetries utilized in our calculations. In section III, we 
present our results for coronene and substituted coronene. 
In the last section we present our conclusions.

\section{Methodology}

Ab-initio DMRG calculations for neutral systems employing molecular 
orbitals have bottlenecks since the calculation of two-electron 
integrals are computationally expensive and consequently these studies employ 
$\sim 50$ active space orbitals \cite{Garnet15}. On the other hand, DMRG calculations with 
localized orbitals have been successfully employed 
for $\pi$-conjugated systems with several hundred $p_z$ orbitals within the PPP Hamiltonian  \cite{Raghu02,Goli16,Sukrit09,Simil13,Mousumi14}.
Ab-initio study of arenes using the DMRG method has also revealed that fully localized orbitals bring about faster 
convergence of energies compared to canonical Hartree-Fock orbitals or split-localized orbitals \cite{Garnet15}, 
making the localized description as the picture of choice.

\subsection{Model Hamiltonian}

We consider the PPP $\pi$-electron Hamiltonian
\cite{Pariser-Parr,Pople} which is a widely employed semi-empirical model 
for studying the behavior of C-based $\pi$-conjugated systems 
\cite{Schulten79-1,Schulten79-2,Schulten87,Schulten86,Soos84,Soos93,Baeriswyl}. 
The PPP Hamiltonian is given by,

\begin{equation}
\begin{split}
\hat H &=\sum_{<i,j>,\sigma} t_0 (\hat c_{i,\sigma}^\dagger \hat c_{j,\sigma}^{} +\mbox{ H.C.})
+ \sum_i \epsilon_i \hat n_i\\
&+\sum_i \frac {U}{2} \hat n_i (\hat n_i-1) + \sum_{i>j} V_{ij}(\hat n_i-z_i)(\hat n_j-z_j)
\end{split}
\label{ppp}
\end{equation}
  
\noindent
In the above $\hat c_{i,\sigma}^\dagger$ ($\hat c_{i,\sigma}^{}$) creates (annihilates) a $\pi$-electron with spin $\sigma$ on
the $p_z$ orbital on C-atom $i$; $\hat c_{i,\sigma}^\dagger \hat c_{i,\sigma}^{}$ is the corresponding 
number operator and $n_i = \sum_{\sigma} c_{i,\sigma}^\dagger \hat c_{i,\sigma}$ is the total number operator.
Here $t_0$ is the transfer integral between bonded C-atoms $i$ and $j$, $\epsilon_i$ is the site energy of the $i$-th C-atom,
$U$ the repulsive Hubbard interaction between two electrons occupying the same $p_z$ orbital, and $V_{ij}$ is the long-range electron 
correlation between C-atoms $i$ and $j$. The latter is obtained from the Ohno interpolation scheme \cite{Ohno,Klopman} (Eq. \ref{ohno}), 

\begin{equation}
V_{ij}=14.397{\left[{\left(\frac{14.397}{U}\right)}^2+r_{ij}^2\right]}^{-\frac{1}{2}}
\label{ohno}
\end{equation}

\noindent
where the distance $r_{ij}$ between the C-atoms i and j are 
in \AA~units while the Hubbard interaction energy term $U$ is in electron-volts (eVs).
Finally, $z_i$ is the local chemical potential, expressed by the number of $\pi$-electrons 
on C-atom $i$ which leaves the site neutral ($z_i=1$ for C-atoms).
In our calculations, we have used standard PPP parameters, $t_0=-2.40$ eV 
and $U=11.26$ eV, that have been widely used over the past several decades {\cite{Salem,Suzuki}}; the employed on-site correlation 
energy $U$ is the sum of the ionization potential and the electron affinity of C in a $sp^2$-hybridized system {\cite{Salem}}. 
Substitution effects can be probed by introducing positive or negative site energies $\epsilon_i$, for mimicking donor and acceptor 
groups respectively, while the site energy for unsubstituted C-atoms is taken as zero. 
 
\subsection{The DMRG Technique}

In the DMRG technique, the full system is divided into two blocks, generally referred to as the left (L) and right (R) blocks, which are 
iteratively grown by a few sites (usually one) at each step. The complete wavefunction of the system is represented in the direct product 
space of the block states. The block space of each individual block is approximated by retaining reduced density matrix 
eigenvectors of the corresponding block with highest eigenvalues and the exponentially growing Hilbert space of a many-body system gets 
well-adapted into a basis space of fixed dimension, independent of the system size. The reduced density matrix of a particular block is 
obtained by employing full system eigenstates while assuming the other block as environment, followed by its diagonalization to obtain 
the block states. Matrices of the block Hamiltonians and of the individual site operators are constructed at the `$l$'-th step in the 
direct product basis of old block states (obtained in the step `$l-1$') and Fock states of the newly added sites. Afterwards, these 
matrices are renormalized employing the new block states constructed at the `$l$'-th step. In the next step, the system is grown by 
adding a few sites to both the left and right blocks and the full system Hamiltonian is constructed in the direct product basis of the block 
states of the left and right blocks along with the Fock states of the newly added sites. The full Hamiltonian matrix is diagonalized to obtain 
targeted eigenstates of the system which are then used to study different physical properties. The above procedure, known as the infinite 
DMRG method, is iteratively repeated until the desired system size is reached. 

Although the infinite DMRG algorithm can be employed to study physical properties at the polymeric limit, for finite-size systems, 
the accuracy of the calculation can be significantly improved by the finite DMRG algorithm. In 
the infinite scheme, the block states at the intermediate steps do not correspond to the final system. This flaw can be resolved 
through construction of block spaces employing wavefunctions corresponding to the final system size. The procedure is termed 
as `sweeping', where iteratively one block is grown at the expense of the other block, while
keeping the final system size fixed. At the final step of a full-sweep, the sizes of the two blocks become $N/2-1$
where, $N$ is the final system size. The finite DMRG procedure is a non-trivial but essential procedure for the 
study of molecular systems as the energies improve significantly following the sweeping procedure. 

For quasi-one dimensional systems, the order in which the new sites are added to build the molecule is important 
to attain high accuracy. For the same DMRG cut-off, different sequences of adding new sites give different energy 
eigenvalues. This is well known in the ab initio DMRG approaches where the order of adding orbitals is determined by 
the entanglement \cite{White99,Garnet02,Garnet03}. 
The order in which the sites are added to build the molecule coronene and its derivative is as follows. At every stage of 
the infinite DMRG iteration, transfer terms are introduced between the sites added earlier (old sites) and the 
new sites as well as between old sites in the two blocks. We add the sites in such a way that the interaction between 
the old site and the new site at any given stage involves as recent an old site as possible. This implies that 
the interaction between the new site and the old site is such that the old site operators have been renormalized 
the fewest number of times possible. In the early DMRG studies of systems with periodic boundary condition, this 
particular requirement is restored by placing one of the newly added sites between the old blocks while placing the 
other new site at the end of a particular old block (L or R).

\begin{figure*}[tp]
\begin{center}
\includegraphics[height=0.8\textheight,width=\textwidth]{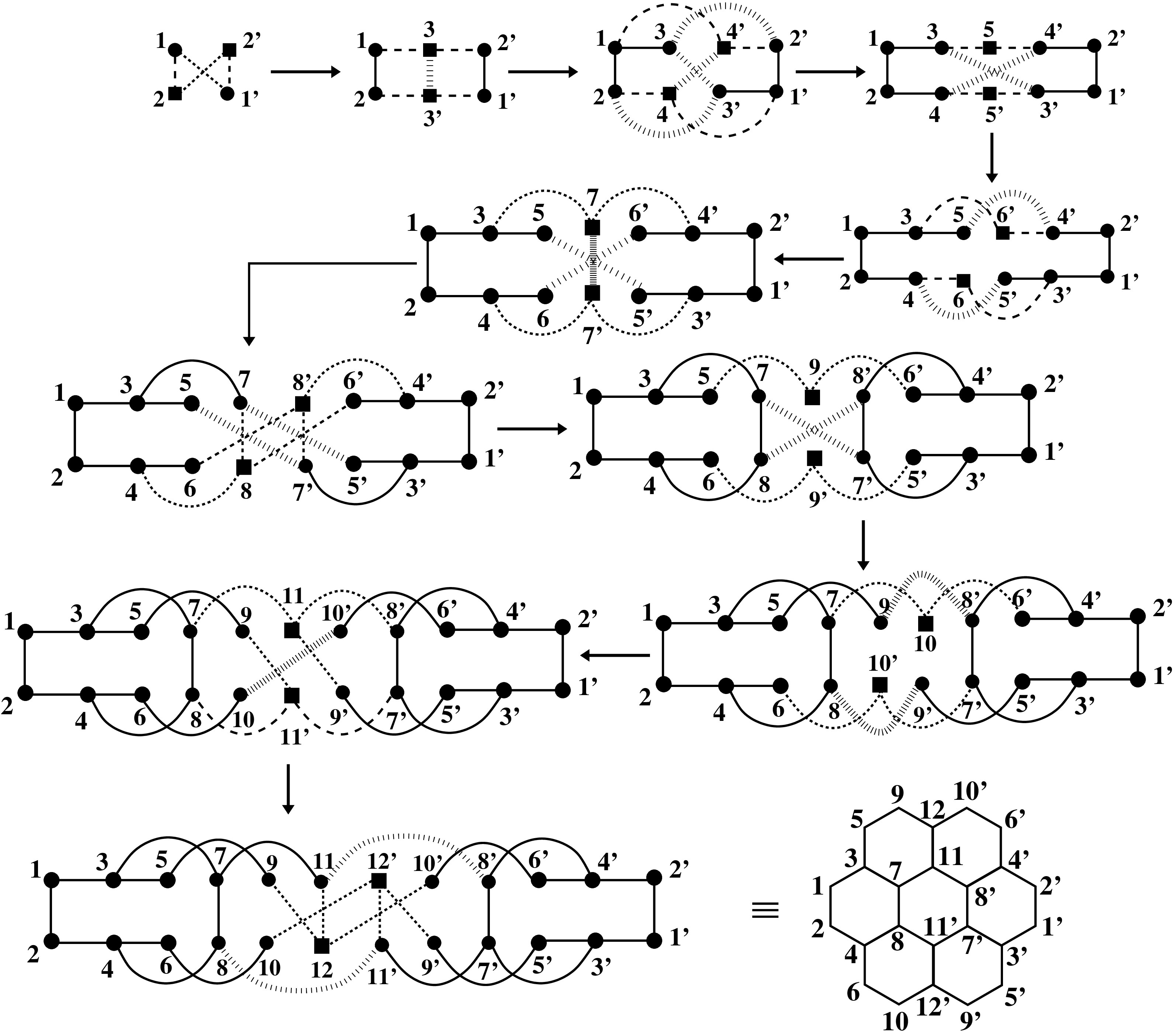}
\caption{\label{block-gen1-6}Construction of coronene molecule in the infinite DMRG method starting from a small 
system (4 sites). The number of connections between the new and the old sites at the intermediate steps are kept 
similar to that in the final system for higher accuracy. At every step of the algorithm two new sites are added, 
one to the system block (L) and other to the environment block (R). The sites in the $L$-block are denoted by 
unprimed numbers while those in the $R$-block are denoted by primed numbers. The newly added sites are denoted 
by filled square ($\blacksquare$) while old sites are denoted by filled circles ($\bullet$). Solid lines are bonds
within a block. The broken lines denote the connections between $\bullet$ and $\blacksquare$. Bonds between the 
two blocks as well as the bond between newly added sites are denoted by hatched lines.}
\end{center}
\end{figure*}

\begin{table}[tp]
\begin{center}
\caption{{\label{gsenergy-6}Ground and lowest optical state energies of coronene
within the non-interacting H\"uckel model, in units of $t_0$, calculated
using the H\"uckel and symmetrized DMRG approaches.}}
\renewcommand{\arraystretch}{1.5}
\begin{ruledtabular}
\begin{tabular}{ccc}
Nature of the state & H\"uckel & Symmetrized DMRG \\
\colrule
Ground state  & $-34.57183$ & $-34.51000$ \\
Optical state & $-33.49345$ & $-33.36570$ \\
\end{tabular}
\end{ruledtabular}
\end{center}
\end{table}

Fig. \ref{block-gen1-6} shows the steps of building the coronene molecule starting from a four-atom ring and adding two new atoms at each 
step. The noninteracting H\"uckel model for this molecule ($\epsilon_i=U=V_{ij}=0$ in Eq.~1) can be solved trivially for ground and 
low-lying excited states. The same can also be obtained using the DMRG algorithm. Comparison of the two results 
(see Table. \ref{gsenergy-6}) provides a stringent check on the DMRG accuracy, since within the noninteracting model, the system block and 
the environment block are more entangled, as compared to interacting models with site-diagonal 
interactions, such as the Hubbard and the PPP models {\cite{Sahoo2012}}. We find that for coronene, 
the ground state energy is accurate to $0.17\%$ and the optically excited state is accurate to $0.38\%$ while the 
optical gap (energy difference between the two) is accurate to $6.1\%$ (exact gap is $1.07838t_0$). Since the accuracy of the 
DMRG method increases with decreasing entanglement, the H\"uckel model provides an upper limit for the errors in the correlated models. 
Additionally, the comparison of 
the DMRG calculations against the H\"uckel model can be employed as an effective tool to calibrate the required block space dimension for desired 
numerical accuracy within these finite-size quasi-one dimensional systems. 

\subsection{Symmetries of the Hamiltonian}

We are particularly interested in the low-energy one and two-photon excited states along with 
the low-lying triplet states of the molecules. However, the number of energy states that 
reside between the ground state and the desired states can be variable and large. Hence, targeting `important' 
states is an almost impossible task without invoking the basic symmetries of the full Hamiltonian. In the present
study, we have utilized the end-to-end interchange symmetry $(C_2)$, $e$-$h$ symmetry and spin-flip 
or parity symmetry $(P)$ of the full system Hamiltonian within the DMRG framework employing a modified algorithm 
for symmetry adaptation \cite{Prodhan18}. In this algorithm, the symmetry operators are expressed as extremely sparse 
matrices, with only one non-zero element per row and column. Consequently, we get rid of the computationally expensive 
Gram-Schmidt orthonormalization procedure during the construction of the symmetry adapted basis states.

The PPP Hamiltonian conserves total spin $(S)$, but it is difficult to adopt total spin conservation within the DMRG 
scheme. In order to target states with different $S$, we exploit the spin-flip symmetry $(P)$ in the $S_z=0$ 
sector, where the Hamiltonian remains invariant as all spins of the system are reversed. The symmetry bifurcates 
the $S_z=0$ space into one subspace with even total spin (designated as `e') and another with odd total spin 
(designated as `o'). In addition, the Hamiltonian for a neutral bipartite system remains invariant under 
$e$-$h$ symmetry, where the creation (annihilation) operator of one sub-lattice is interchanged by 
annihilation (creation) operator, while in the other sub-lattice, the interchange accompanies a phase factor of 
$-1$. The eigenstates can be labeled `+' or `-' depending upon the eigenvalue ($+1$ or $-1$) while operated by 
$e$-$h$ symmetry operator. Finally, the full system eigenstates can be labeled $A$ or $B$, based on their parity 
(even or odd) with respect to $C_2$ operation. 

The three symmetry operators and their products along with identity form an Abelian group which sub-divides 
the $S_z=0$ space into eight subspaces. In general, the ground state has even character with respect to every 
symmetry operation and lies in the ${}^eA^+$ subspace. Optical one-photon states remain in ${}^eB^-$ space
while the two-photon states have the same symmetry characters as the ground state. 
The lowest triplet state energy is in the $S_z=1$ space where the $P$ symmetry cannot be employed and it remains in the $B^+$ 
space.

In each symmetry subspace, we have calculated a few low-lying eigenstates of the Hamiltonian to ascertain the spectra in the low 
energy region. However, for the calculation of the transition dipole moments, the average reduced density matrix is employed 
instead of the reduced density matrix corresponding to a single state, in order to attain a common block space description.
The average reduced density matrix \cite{White93} is defined by $\rho=\sum_{i}\omega_{i}\rho_{i}$ where $\rho_i$
is the reduced density matrix corresponding to eigenstates $|i \rangle$. $\omega_i$ are the 
weights of the corresponding eigenstates, which we have taken as $\omega_i=1/W$, where $W$ is the number of 
low-lying eigenstates computed in the symmetry subspace. The block states obtained from the average reduced 
density matrix are employed for the DMRG calculations. The magnitude of the cut-off in the block space dimension does not 
lead to admixture of the different symmetry states in the state-average DMRG calculations, since we have retained all symmetry partners 
of the block states in our algorithm.

\subsection{Two-Photon Absorption Cross-Section}

Two-photon absorption (TPA) is a third order nonlinear optical process which involves simultaneous absorption of two 
photons and is related to the imaginary part of the second order hyperpolarizability 
$\chi^{(3)}(-\omega;\omega,-\omega,\omega)$, where $\hbar\omega$ is half of the excitation energy of the two-photon 
state $|TP\rangle$ $(\hbar \omega=(E_{TP}-E_G)/2)$. 
We have employed the correction vector (CV) technique for computing the TPA cross-section.
The first order CV is calculated employing the inhomogeneous linear algebraic equation, 
\begin{equation}
(\hat H-E_G+\hbar \omega )|\phi_i^{(1)}(\omega)\rangle = \tilde{\mu}_i |G\rangle
\label{cv}
\end{equation}

\noindent
where $H$ is the Hamiltonian matrix, $|G\rangle$ is the ground state with energy $E_G$ and $\tilde{\mu}_i$ is the 
$i$-th component of the dipole displacement operator, $\tilde{\mu}_i=\hat{\mu}_i-\langle G|\hat{\mu}_i|G \rangle$ 
$(i=x,y,z)$. The linear algebraic equations are solved efficiently
employing the small matrix algorithm developed by Ramasesha \cite{Ramasesh90}. On expansion in the basis of the 
excited states $\{|R\rangle\}$, the correction vector can be written as,
\begin{equation}
|\phi_i^{(1)}(\omega)\rangle = \sum_R \frac{\langle R|\tilde{\mu}_i|G\rangle}{E_R-E_G+\hbar\omega} |R\rangle
\end{equation}  

The expression for the $ij$-th element of the two-photon transition matrix is given by \cite{Agren2000},
\begin{equation}
\begin{split}
S_{ij}(\omega)&=\sum_R \left[ \frac{\langle G|\tilde{\mu}_i|R\rangle \langle R|\tilde{\mu}_j|TP\rangle}{E_R-E_G-\hbar\omega}
+\frac{\langle G|\tilde{\mu}_j|R\rangle \langle R|\tilde{\mu}_i|TP\rangle}{E_R-E_G-\hbar\omega}\right] \\
&=\langle \phi_i^{(1)}(-\omega)|\tilde{\mu}_i|TP\rangle + \langle \phi_j^{(1)}(-\omega)|\tilde{\mu}_i|TP\rangle\\
\end{split}
\end{equation}

\noindent
while, the TPA cross-section for a linearly polarized monochromatic light of a randomly oriented sample as in 
solution or gas phase is given by \cite{Agren2000},
\begin{equation}
\delta_{TPA}=\frac{1}{15} \sum_{i,j=x,y,z}(S_{ii}S^{\ast}_{jj}+2S_{ij}S^{\ast}_{ij})
\label{tpa}
\end{equation}

\section{Computational Results}

\begin{figure}[t]
\begin{center}
\includegraphics[height=5.5cm]{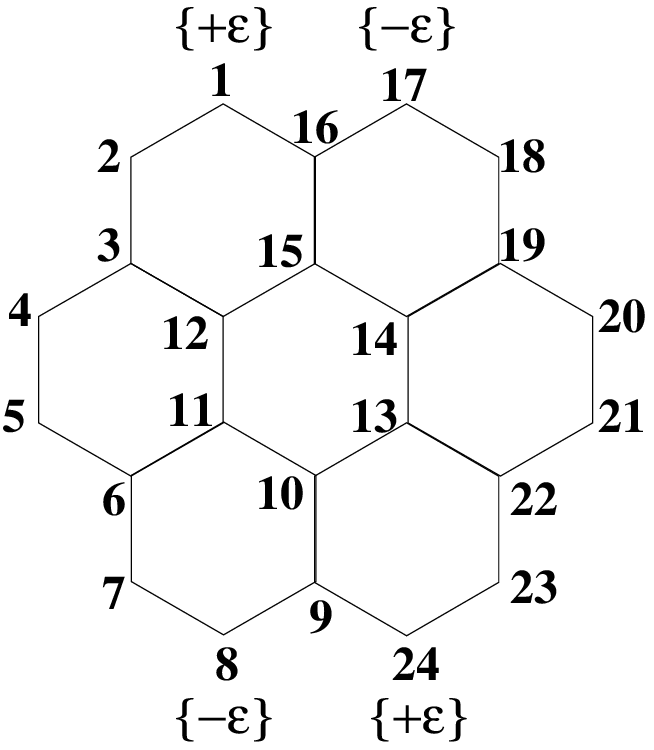}
\caption{\label{molecule}(Color online) Schematic diagram of coronene molecule. The sites of substitution in substituted coronene 
are also indicated; $+\epsilon$ represents a donor site while $-\epsilon$ represents an acceptor site.}
\end{center}
\end{figure}

We have studied a few low-lying states of coronene in different symmetry subspaces
within the PPP model and the relevant energy gaps, defined as differences from the 
ground state, are tabulated in Table \ref{energy-ppp1}. We have also calculated 
energies in the corresponding symmetry subspaces for the substituted coronene of 
Fig. \ref{molecule} (see Table. \ref{energy-ppp2}), with donor and acceptor groups of 
equal strength ($|\epsilon|=1.0$ eV). For all our calculations, we maintained 
truncated block space dimension of $\sim 1000$. 
We have used a single spatial symmetry, the $C_2$ symmetry whose axis is perpendicular 
to the molecular plane, as our DMRG calculation 
cannot handle more than one symmetry axis. 
We label all the eigenstates using the $D_{2h}$ subgroup symmetry, which is simpler than
determining the irreducible representations of the $D_{6h}$ point group symmetry.
In the $D_{6h}$ point group representation, allowed optical transitions 
from the ground state with $A_{1g}$ symmetry are only to doubly degenerate $E_{1u}$ states 
whose transition dipoles lie in the molecular plane. In the $D_{2h}$ subgroup, these states 
remain as two-fold degenerate $B_{2u}$ and $B_{3u}$ states respectively, but now their transition dipoles 
lie strictly along orthogonal y- and x-axes respectively. Consequently, use of $D_{2h}$ symmetry subgroup
instead of $D_{6h}$ simply implies that the doubly degenerate optical states, with mixed polarizations
in the molecular plane, are assumed to have distinct polarizations along the Cartesian axes and \textit{no
information has been lost}. Within the DMRG calculations of these doubly degenerate states, however, calculated
transition dipoles along any one direction, picks up a weak orthogonal component even when $D_{2h}$ symmetry is employed.
The lowest optical states obtained may not be the states with highest transition dipole moment, which will be prominent 
in UV-visible spectroscopy but corresponds to the lowest energy state of the appropriate symmetry subspace \cite{Mazumdar14-1}.

\subsection{Correlation Strength and Ordering of Excited States}

As has been shown explicitly in calculations of linear polyenes \cite{Hudson72,Schulten72,Hudson82,Ramasesha84-2}, 
the most important consequence of 
strong electron correlations is the energy ordering of excited states according to their {\it ionicities}. In the language
of valence bond (VB) theory, eigenstates are covalent if they are dominated by VB diagrams in which the $p_z$ orbitals of C-atoms
are neutral, {\it i.e.}, singly occupied. Similarly, eigenstates are ionic if in the dominant VB diagrams, 
there occur at least one pair of $p_z$ orbitals that are positively and negatively charged. One simple measure 
of the ionicity within the PPP Hamiltonian of Eq.~(1), is the expectation value of $\langle n_{i,\uparrow}  n_{i,\downarrow}\rangle$,
which measures the probability of double occupancy in the $p_z$ orbital at site $i$.
Within the noninteracting H\"uckel model, $\langle n_{i,\uparrow}  n_{i,\downarrow}\rangle$ is exactly 0.25 for the ground state.
The ground state of the interacting Hamiltonian is more covalent than that of the noninteracting Hamiltonian and 
$\langle n_{i,\uparrow}  n_{i,\downarrow}\rangle < 0.25$ for the former. Experimentally, the ionicities of the excited states is a 
more relevant quantity. Within the noninteracting model, there is no one-to-one correspondence between energies of states and their
ionicities. Within the interacting model, however, predominantly covalent states occur energetically below predominantly ionic states.

\begin{table*}[t]
\small
\begin{center}
%\begin{threeparttable}
\caption{{\label{energy-ppp1} Energies of low-lying two-photon states, optical states, triplet states and a few other optically dark states 
in coronene, relative to the ground state. Although coronene has $D_{6h}$ symmetry, here the states are labeled by the symmetry representations of its subgroup $D_{2h}$. Whenever a state cannot be uniquely labeled due to lower symmetry employed in the study, both possible labels for the state have been given. The transition dipole moment (in Debye) from ground state to the excited states $(\mu_{tr,x/y})$ along specific axes are also specified in the last two columns. Energies determined by the UV-visible spectroscopy are also mentioned in the footnotes (a) and (c).}}
\renewcommand{\arraystretch}{1.5}
%\begin{tabular}{|>{\centering\arraybackslash}p{3.5cm}|*{4}{>{\centering\arraybackslash}p{2.50cm}|}}
\begin{ruledtabular}
%\begin{threeparttable}
\begin{tabular}{ccccc}
\multicolumn{5}{c}{Coronene} \\
\colrule
Nature of the state & State & Energy Gap (eV) & $\mu_{tr,x}(D)$ & $\mu_{tr,y}(D)$ \\
\colrule
Two-photon & $(2\thinspace{}^{1}A_{g}^{+}/ 1\thinspace{}^{1}B_{1g}^{+})$ & $3.97$ & $0.00$ & $0.00$ \\
           & $(1\thinspace{}^{1}B_{1g}^{+}/ 2\thinspace{}^{1}A_{g}^{+})$ & $4.09$ & $0.00$ & $0.00$ \\
           & $(3\thinspace{}^{1}A_{g}^{+}/ 2\thinspace{}^{1}B_{1g}^{+})$ & $5.08$ & $0.00$ & $0.00$ \\
\colrule
Optical & $1\thinspace{}^{1}B_{2u}^{-}$ & $4.83$ \footnotemark & $0.81$\footnotemark & $8.43$ \\
\footnotetext[1]{$4.10$~eV\cite{Mazumdar14-1}; $4.06-4.27$~eV\cite{Clar}; $4.21$~eV\cite{Schmidt77};
$4.07-4.23$~eV\cite{Patterson42}; $4.12-4.44$~eV\cite{Tanaka65}; $4.06-4.27$~eV\cite{Stephens68};
$4.28$~eV\cite{Abouaf09}; $4.06$~eV\cite{Aihara70}; $4.09$~eV\cite{Ohno72}; $4.27$~eV\cite{Ehrenfreund92}.}
\footnotetext[2]{Non-zero value of transition dipole moment along polarization direction forbidden by
symmetry is an artifact of the calculations as average density matrices, calculated from eigenstates of different
symmetry subspaces, are employed to determine the transition dipole moment. However, the errors are negligible as
intensities depend on the square of the transition dipole moment.}
        & $1\thinspace{}^{1}B_{3u}^{-}$ & $4.87$ \footnotemark[1] & $7.51$ & $0.49$\footnotemark[2] \\
\colrule
Triplet & $(1\thinspace{}^{3}B_{2u}^{+}/ 1\thinspace{}^{3}B_{3u}^{+})$ & $2.35$ \footnotemark & $0.00$ & $0.00$ \\
\footnotetext[3]{$2.40$~eV\cite{Ohno72,Abouaf09}.}
        & $(1\thinspace{}^{3}B_{3u}^{+}/ 1\thinspace{}^{3}B_{2u}^{+})$ & $3.00$ & $0.00$ & $0.00$ \\
        & $(2\thinspace{}^{3}B_{2u}^{+}/ 2\thinspace{}^{3}B_{3u}^{+})$ & $3.02$ & $0.00$ & $0.00$ \\
        & $(1\thinspace{}^{3}A_{g}^{+}/  1\thinspace{}^{3}B_{1g}^{+})$ & $3.35$ & $0.00$ & $0.00$ \\
\colrule
Dark states & $(1\thinspace{}^{1}B_{2u}^{+}/ 1\thinspace{}^{1}B_{3u}^{+})$ & $2.82$ & $0.00$ & $0.00$ \\
%            & $n\thinspace{}^{1}B_{2u}^{-}$/ $n\thinspace{}^{1}B_{3u}^{-}\tnote{4}$ & $3.88$ & $0.28$ & $1.74$  \\
            & $(1\thinspace{}^{1}A_{g}^{-}/ 1\thinspace{}^{1}B_{1g}^{-})$ & $3.88$ & $0.28$ & $1.74$  \\
            & $(1\thinspace{}^{1}B_{1g}^{-}/ 1\thinspace{}^{1}A_{g}^{-})$ & $4.73$ & $0.00$ & $0.00$ \\

\end{tabular}
%%%\begin{tablenotes}
%%%\item[1]{$4.10$~eV\cite{Mazumdar14-1}; $4.06-4.27$~eV\cite{Clar}; $4.21$~eV\cite{Schmidt77}; 
%%%$4.07-4.23$~eV\cite{Patterson42}; $4.12-4.44$~eV\cite{Tanaka65}; $4.06-4.27$~eV\cite{Stephens68}; 
%%%$4.28$~eV\cite{Abouaf09}; $4.06$~eV\cite{Aihara70}; $4.09$~eV\cite{Ohno72}; $4.27$~eV\cite{Ehrenfreund92}.}\\
%%%\item[2]{Non-zero value of transition dipole moment along polarization direction forbidden by
%%%symmetry is an artifact of the calculations as average density matrices, calculated from eigenstates of different
%%%symmetry subspaces, are employed to determine the transition dipole moment. However, the errors are negligible as
%%%intensities depend on the square of the transition dipole moment.}\\
%%%\item[3]{$2.40$~eV\cite{Ohno72,Abouaf09}.}\\
%\item[4]{n implies that the level numbering is unknown. This state has very small intensity compared to the other two
%optical states and hence, classified as dark state.}
%%%\end{tablenotes}
%\end{threeparttable}
\end{ruledtabular}
\end{center}
\end{table*}

Fortunately, this relative ordering, lowest covalent excited states occurring below lowest ionic states, can be tested optically
in centrosymmetric systems with distinct one-photon and two-photon states. As was recognized long back, dipole selection rules in centrosymmetric
systems dictate that transition dipole matrix elements are nonzero only between states with opposite parity and $e$-$h$ 
symmetries \cite{Hudson82,Baeriswyl92}. In the context of neutral $\pi$-conjugated molecules, this
means that one-photon transition from the even parity $eA^+$ covalent ground state can occur only to odd parity $eB^-$ states. Note that, 
this also implies that the lowest two-photon states, which are dipole-coupled to the $eB^-$ states are covalent (see Eq.~(5)), and are hence
likely to occur below the lowest one-photon state. The occurrence of the lowest two-photon state below the lowest one-photon 
state has been experimentally confirmed in linear polyenes \cite{Hudson72}.

The theoretical measure of the correlation strength for a charge-neutral C-based molecule
to a first approximation is the ratio of the effective on-site correlation, to the width of the one-electron energy spectrum.
The effective onsite correlation within the PPP model is $U_{PPP}\sim (U-V_{12})$, where $V_{12}$ is the nearest-neighbor Coulomb interaction.
This quantity is independent of dimensionality. In contrast, the width of the one-electron energy spectrum 
increases with dimensionality, implying therefore that the effective correlation strength is smaller in two dimension that in one dimension.
The relative energies of the lowest one versus two-photon states in graphene nanofragments can therefore not be guessed based on the known 
results for polyenes.

Aside from its covalent nature, another aspect of the lowest two-photon state in linear polyenes has been of interest, viz., its
characterization as a bound state of the two lowest triplet excitations T$_1$ of the polyene \cite{Schulten87,Ramasesha84-2}. 
Indeed, several of the lowest two-photon
states in linear polyenes are superpositions of triplet excitations T$_1$ as well as higher energy T$_2$, T$_3$ etc. 
(the two triplets that constitute an excited covalent singlet need not be identical) \cite{Schulten87}. 
In other words, even parity covalent excited states in linear polyenes are necessarily superpositions of two triplets. It is not 
a priori clear that this
will hold true in polycyclic aromatic hydrocarbons such as coronene, where there occur C-atoms that are different, peripheral versus internal,
as well as bi-coordinate versus tri-coordinate.  

In Table \ref{energy-ppp1}, we have listed the energies of the lowest optical states, the lowest triplet states and the lowest two-photon states,
relative to the ground state. We have also included the lowest states that are optically dark under both one- and two-photon 
excitation. The latter are equivalent to the covalent $B_u^+$ states of polyenes {\cite{Ramasesha84-2}}, each of which
can be viewed as superposition of covalent $B_u$ triplets $T_i$ and $T_j$, $i\ne j$. 
Because of our use of a single C$_2$ symmetry element,
we are unable to distinguish between $A_g^+$ and $B_{1g}^+$ states that are degenerate in the noninteracting limit but are
nondegenerate within the correlated PPP Hamiltonian \cite{Mazumdar14-1}. 
The $^1$B$_{2u}^-$ and  $^1$B$_{3u}^-$ states are degenerate, but
the triplet states ${}^3B_{2u}^+$ and ${}^3B_{3u}^+$ need not to be degenerate in coronene \cite{Mazumdar14-1}.
We have therefore assigned the triplet at $2.35$ eV to ${1}^3B_{2u}^+$/${1}^3B_{3u}^+$, while the one at $3.0$ eV is assigned to ${1}^3B_{3u}^+$/${1}^3B_{2u}^+$.
We have given similar labels to the lowest two-photon states, where two multiple assignments are possible.

\begin{table*}[tp]
\begin{center}
\caption{{\label{energy-ppp2} Energies of low-lying states states in substituted coronene  are 
tabulated below. The transition dipole moment (in Debye) from ground state to the excited states $(\mu_{tr,x/y})$ 
along specific axes are also specified in the last two columns.}}
\renewcommand{\arraystretch}{1.5}
%\begin{tabular}{|>{\centering\arraybackslash}p{3.5cm}|*{4}{>{\centering\arraybackslash}p{2.0cm}|}}
\begin{ruledtabular}
\begin{tabular}{ccccc}
\multicolumn{5}{c}{Substituted coronene} \\
\colrule
Nature of the state & State & Energy Gap (eV) & $\mu_{tr,x}(D)$ & $\mu_{tr,y}(D)$ \\
\colrule
Two-photon & $2\thinspace{}^{1}A_{g}$ & $4.01$ & $0.00$ & $0.00$ \\ 
           & $3\thinspace{}^{1}A_{g}$ & $4.81$ & $0.00$ & $0.00$ \\
	   & $4\thinspace{}^{1}A_{g}$ & $4.90$ & $0.00$ & $0.00$ \\
\colrule
Optical & $1\thinspace{}^{1}B_{u}$ & $2.90$ & $0.00$ & $0.23$ \\
        & $2\thinspace{}^{1}B_{u}$ & $3.87$ & $0.12$ & $1.46$ \\
        & $3\thinspace{}^{1}B_{u}$ & $5.18$ & $0.17$ & $7.52$ \\
        & $4\thinspace{}^{1}B_{u}$ & $5.97$ & $4.10$ & $0.59$ \\
\colrule
Triplet & $1\thinspace{}^{3}B_{u}$ & $2.29$ & $0.00$ & $0.00$ \\
        & $2\thinspace{}^{3}B_{u}$ & $2.92$ & $0.00$ & $0.00$ \\
        & $3\thinspace{}^{3}B_{u}$ & $3.48$ & $0.00$ & $0.00$ \\
\end{tabular}
\end{ruledtabular}
\end{center}
\end{table*}

The lowest two-photon energy along with the lowest triplet energy calculated in coronene are in 
excellent agreement with the previously reported values by Aryanpour et al., obtained using the MRSDCI technique \cite{Mazumdar14-1}.
The calculated energy of the lowest two-photon state (that occurs in the ${}^1A_g^+$ subspace) reported earlier is $3.96$ eV for coronene, 
to be compared against $3.97$ eV
found in our calculations. These numbers match very well against the experimental two-photon absorption spectrum \cite{Mazumdar14-1}.
The earlier reported lowest triplet energy is $2.38$ eV (experimentally
identified peak is at $\sim 2.40$ eV, probed by phosphorescence and electron energy loss
spectroscopy \cite{Ohno72,Abouaf09}), which is also close to $2.35$ eV obtained by us. However, the lowest one-photon energy and 
relative position of the lowest optical state with respect to the lowest two-photon state do not agree well 
with the earlier study \cite{Mazumdar14-1,Mazumdar14-2}. In coronene, the energy of the lowest optical $1\thinspace{}^{1}B_{2u}^{-}$ and
$1\thinspace{}^{1}B_{3u}^{-}$ states calculated earlier \cite{Mazumdar14-1,Mazumdar14-2} is $4.11$ eV while we have found two nearly degenerate 
excitations at higher energies $4.83$ and $4.87$ eV, respectively. 

The experimental linear absorption spectrum of coronene shows a prominent 
absorption band in this region with maximum at $4.06-4.30$ eV both in 
solution \cite{Mazumdar14-1,Clar,Schmidt77,Nijegorodov01,Patterson42,Tanaka65,Stephens68} and vapor 
phase \cite{Aihara70,Ohno72,Ehrenfreund92,Abouaf09}.
The discrepancy between our result and the previous computational result  arises from the use of ``bare'' PPP-Ohno parameters in the present 
work as opposed to ``screened'' parameters
in the previous work \cite{Mazumdar14-1,Mazumdar14-2}. 

\begin{figure}[b]
\begin{center}
\includegraphics*[height=5.5cm]{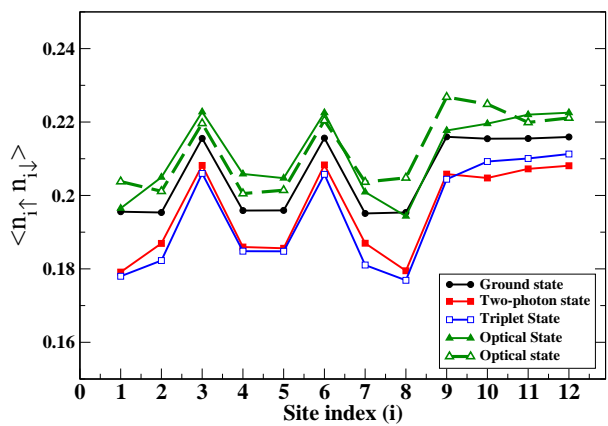}
\caption{\label{dblocc}(Color online) The probability of double occupancy of C-atoms by electrons plotted against site index 
(see Fig. \ref{molecule}) for coronene. Sites related by $C_2$ symmetry are perfectly equivalent and hence are not shown. Lines are guides 
to the eye only. The two different plots for the optical states correspond to the two nearly degenerate states.}
\end{center}
\end{figure}

We see in Table \ref{energy-ppp1} a low-lying state at an energy of $3.88$ eV, with a small transition dipole moment.
This state was found to be optically forbidden in the MRSDCI calculations \cite{Mazumdar14-1,Mazumdar14-2}.
The weak dipole coupling found in the present calculations probably results from our incorporation of only one $C_2$ symmetry 
axis or our use of the  average density matrices, obtained from eigenstates of different symmetry subspaces, that can lead to weak spatial 
symmetry violation. Indeed, we revisited the calculations without employing the $C_2$ symmetry and found the energy of this 
state remains unchanged. However, this state lies in the `-' subspace of $e$-$h$ symmetry with a small transition dipole moment. Therefore,
we conclude that this state corresponds to the $1\thinspace{}^{1}A_{g}^{-}/ 1\thinspace{}^{1}B_{1g}^{-}$ state and in agreement with the 
previous study {\cite{Mazumdar14-1,Mazumdar14-2}}.  
The calculated small transition dipole moment is thus an artifact, although weak violation of $e$-$h$ symmetry, as would 
occur in the real molecule, can lead to observable absorption. 
Indeed, as pointed out in the earlier theory-experiment work \cite{Mazumdar14-1}, this ``forbidden'' state is seen as a weak absorption 
experimentally.

The $2.82$ eV excitation in coronene, on the other hand, is strictly forbidden 
as it belongs to the same $e$-$h$ symmetry subspace as the ground state. This argument is supported by the fact 
that this state acquires some intensity on breaking the $e$-$h$ symmetry by introducing substituents, as can be 
seen from Table. \ref{energy-ppp2}. 
Appearance of absorption peak at $\sim 2.95$ eV in thin films of coronene also suggest presence of a singlet 
state close to the lowest triplet state {\cite{Kropp67,Dutta98}}. 

Substitution by donor-acceptor groups does not seem to have appreciable effect on the energy gaps between the states, 
although the lifting of e-h symmetry allows optical transitions to states which are 
dipole-forbidden in the unsubstituted molecule. The extent of mixing of different symmetry states of the 
unsubstituted system due to substitution depends on the strength of the donor-acceptor groups. 

The spin gap (energy difference between lowest triplet state and ground state) is also a good measure of the effective correlation strength. 
Stronger the effective correlation, the smaller is the spin gap. Based on the similar spin gaps in the unsubstituted versus substituted 
molecules, we conclude that the effective correlation strengths are the same in the two molecules. This indicates that the
previous claim of stronger correlation effect in lower symmetry molecules \cite{Mazumdar14-2} may be an oversimplification.

In Fig. \ref{dblocc} we have plotted the $\langle n_{i,\uparrow}n_{i, \downarrow} \rangle$ for each of the C-atoms in coronene, for the ground state, the optical 
$1\thinspace{}^1B_{2u}^{-}$ and $1\thinspace{}^1B_{3u}^{-}$ states, the lowest two-photon state at $3.97$ eV, and the lowest triplet state. As expected,
$\langle n_{i,\uparrow}n_{i, \downarrow} \rangle$ for the ground state is smaller than $0.25$ for all C-atoms, indicating its covalent character.
The same expectation value is larger for the optical state, also as anticipated for this ionic state. Interestingly, $\langle n_{i,\uparrow}n_{i, \downarrow} \rangle$
for the lowest triplet and the lowest two-photon state are both {\it smaller} than that of the ground state, indicating, (i) covalent 
character larger than that of the ground state, and (ii) nearly equal covalent character in both. The equality between the lowest
triplet and the lowest two-photon state is surprising, given that the latter is not a simple two-triplet state.

\subsection{TPA cross-section}

\begin{table*}[t]
\footnotesize
\begin{center}
\caption{\label{two-photon} Two-photon transition matrix elements along with TPA cross-section for lowest 
two-photon states in unsubstituted coronene molecule. Possible 
symmetry labels are provided wherever unique symmetry label cannot be determined. Transition matrix elements as 
well as TPA cross-sections are given in atomic units.}
\begin{ruledtabular}
%\begin{tabular}{ccdddd}
\renewcommand{\arraystretch}{1.5}
\begin{tabular}{cccccc}
& Two-photon state & \multicolumn{1}{l}{~~~$S_{xx}$} & \multicolumn{1}{l}{~~~$S_{yy}$} & \multicolumn{1}{l}{~~~$S_{xy}$} & \multicolumn{1}{l}{$\delta_{TPA}$} \\
\colrule
         & $(2\thinspace{}^{1}A_{g}^{+}/ 1\thinspace{}^{1}B_{1g}^{+})$ & $10.68$ & $-67.05$ & $-0.97$ & $821.56$  \\
Coronene & $(1\thinspace{}^{1}B_{1g}^{+}/ 2\thinspace{}^{1}A_{g}^{+})$ & $1.94$ & $-0.97$ & $-35.95$ & $349.39$ \\
         & $(3\thinspace{}^{1}A_{g}^{+}/ 2\thinspace{}^{1}B_{1g}^{+})$ & $23.32$ & $-69.96$ & $1.94$ & $868.77$ \\
\end{tabular}
\end{ruledtabular}
\end{center}
\end{table*}

In Table. \ref{two-photon}, we have tabulated the TPA cross-sections for low-lying two-photon states in coronene along with two-photon transition matrix elements. In coronene, we find that the higher energy two-photon 
state $3\thinspace{}^{1}A_{g}^{+}$/$2\thinspace{}^{1}B_{1g}^{+}$ 
has larger TPA cross-section than that of the lowest two-photon states.
This theoretical result is in qualitative agreement with the experimental solution two-photon measurements in this energy region.
Since the two Cartesian axes are equivalent in coronene, the transition matrix elements $S_{xx}$ and $S_{yy}$
should be nearly same. However, in our calculations, we found $|S_{yy}|>|S_{xx}|$ in most of the cases, which we 
attribute to the fact that we have used only one $C_2$ symmetry axis while targeting the states as well as to the 
use of average density matrices. When the states are nearly degenerate, these approximations could break the true 
symmetry which the eigenstates will otherwise possess. 

\section{Discussion and Conclusion}

We have studied the lowest energy states and their relative orderings in two finite centrosymmetric graphene nanoflakes
within the PPP $\pi$-electron Hamiltonian, using the DMRG approach. 
Electron correlations drive covalency in both molecules and also change the relative orderings of one- versus
two-photon excitations. As in linear polyenes, the lowest triplet and the lowest two-photon states are
covalent in the language of VB theory, and occur below the lowest one-photon optical state. The proximity 
in energy between the ionic one-photon state and the covalent two-photon state, relative to that in the polyenes,
however is an indication of relatively weaker correlation effect in these two-dimensional molecules with wider one-electron
energy spectrum. Additionally, the lowest two-photon state is not a simple two-triplet state, unlike in the polyenes.
We believe that this is a consequence of different topology in these polycyclic hydrocarbons, in which there occur
C-atoms with both two and three nearest neighbors. The occurrence of higher energy two-photon states that are
two-triplets \cite{Mazumdar14-2} indicates that in these two-dimensional molecules there occur two different kinds
of covalent states, which may or may not be simply classified as two-triplet. This relationship between the natures of
covalent states with topology, along with the correlation effects in graphene fragments of larger and larger size, are topics of
ongoing and future interest. 
%
%Effective electronic correlation strength in higher-dimensional system is a very subtle parameter which is 
%not only controlled by the one-particle bandwidth but also by the topology of the system. Indeed, correlation 
%plays a crucial part in the relative ordering of the one-photon, two-photon and triplet states. In this 
%article, we have studied the lowest energy level ordering in some polyaromatic systems, which mimics graphene 
%nanoflakes, using the PPP model with standard parameters for $sp^2$ carbon atoms employing symmetry-adapted DMRG 
%technique. We have obtained energy gaps for several one-photon, two-photon and triplet states in coronene, ovalene
%and substituted coronene and found that the energy level ordering in these molecules strongly depend on the 
%topology of the transfer part of the Hamiltonian besides the interaction parameters. The effective strength of 
%correlation in these molecules
%can be inferred from the energy level ordering and their energy gaps from the ground state. As these molecules 
%are smaller versions of graphene, these studies will shed light on how correlation effect can be tuned and 
%utilized within this two-dimensional network for future nanoelectronic applications. It is also of interest to 
%determine how the spectrum evolves with increasing size of these graphene dots. This is a subject for future 
%research.

%\authorcontributions{Investigation, Suryoday Prodhan, Sumit Mazumdar and S. Ramasesha; Funding acquisition, Sumit Mazumdar and S. Ramasesha;
%Writing – original draft, Suryoday Prodhan, Sumit Mazumdar and S. Ramasesha.}

\begin{acknowledgments}
The authors are grateful for financial support from the Indo-US Science and            
Technology Forum that was instrumental in the creation of a Joint Center, which made this collaborative research possible.
SR is thankful to the Department of Science and Technology, India for financial support.
SM acknowledges partial support from U. S. NSF grants CHE-1764152 and the UA-REN Faculty Exploratory Research Grant.
SP acknowledges CSIR India for a senior research fellowship.
\end{acknowledgments}

%\conflictsofinterest{The authors declare no conflict of interest.}

%\reftitle{References}
%\bibliographystyle{biochem}
%\externalbibliography{yes}
\bibliography{manuscript_final_2}

\end{document}